\documentclass[runningheads]{llncs}

\usepackage{graphicx} 
\usepackage{hyperref}
\usepackage{todonotes}
\usepackage{soul}
\usepackage{xspace}
\usepackage{xcolor}
\usepackage{amsmath}
\usepackage{amsfonts}
\usepackage{algorithm}
\usepackage{algpseudocode}
\usepackage[capitalize]{cleveref}

\newcommand{\val}{\ensuremath{\mathcal{V}}}

\newcommand{\ledger}{\ensuremath{\mathcal{L}}}
\newcommand{\tx}{\ensuremath{\textsf{tx}}}
\newcommand{\fullnode}{\ensuremath{\textsf{F}_h}\xspace}
\newcommand{\fee}{\ensuremath{\textsf{fee}}\xspace}
\newcommand{\reward}{\ensuremath{\mathcal{F}}\xspace}
\newcommand{\maxfee}{\ensuremath{\bar{\mathcal{F}}}}
\newcommand{\A}{\ensuremath{\mathcal{A}}\xspace}

\newcommand{\prot}{\ensuremath{\Pi}\xspace}
\newcommand{\stake}{\ensuremath{D}\xspace}
\newcommand{\mempool}{\ensuremath{\mathcal{M}}}
\newcommand{\budget}{\ensuremath{\mathcal{B}}}
\newcommand{\qc}{\ensuremath{\mathsf{QC}}}
\newcommand{\cens}{\ensuremath{\mathsf{cens}}}
\newcommand{\bw}{\ensuremath{\rho}}

\newcommand{\Wprod}{\ensuremath{\mathsf{W}}\xspace}
\newcommand{\Wcons}{\ensuremath{\mathsf{C}}\xspace}
\newcommand{\seen}{\ensuremath{\mathcal{S}}}
\newcommand{\out}{\ensuremath{\widehat{\ledger}}}

\newcommand{\comm}{\ensuremath{\mathcal{C}}}
\newcommand{\bwin}{\ensuremath{u}\xspace}
\newcommand{\handover}{\ensuremath{\mathsf{HO}}}

\newcommand{\fn}[1]{\texttt{#1}}
\newcommand{\var}[1]{\ensuremath{\textsf{#1}}}

\newcommand{\receive}{\fn{receive}}
\newcommand{\multicast}{\fn{multicast}}
\newcommand{\schedule}{\fn{schedule}}
\newcommand{\request}{\fn{request}}

\newcommand{\bootstrap}{\fn{bootstrap}}
\newcommand{\signers}{\fn{signers}}
\newcommand{\committee}{\fn{committee}}

\newcommand{\quorumstore}{\var{quorum\_store}}

\newcommand{\return}{\State\Return}

\algblock{On}{EndOn}
\algrenewtext{On}[1]{\textbf{upon} #1}
\algrenewtext{EndOn}{\textbf{end upon}}

\algrenewcommand\textproc{\texttt}

\newenvironment{proofof}[1]{\par\addvspace{6pt}\noindent\textit{Proof of #1.}\ }{\qed\par\addvspace{6pt}}
\let\llncsendproof\endproof
\renewcommand{\endproof}{\qed\llncsendproof}

\newcommand{\com}[1]{}

\makeatletter
\renewcommand{\theHALG@line}{\thealgorithm.\arabic{ALG@line}}
\renewcommand{\theALG@line}{\thealgorithm.\arabic{ALG@line}}
\makeatother

\crefname{proposition}{Proposition}{Propositions}
\Crefname{proposition}{Proposition}{Propositions}

\crefname{ALC@line}{line}{lines}
\Crefname{ALC@line}{Line}{Lines}

\newif\ifanonymous

\anonymousfalse
\title{Playing the SMR Game with Existential Honesty}
\titlerunning{Playing the SMR Game with Existential Honesty}
\ifanonymous
  \author{}
  \institute{}
\else
  \author{Zeta Avarikioti\inst{1,2} \and
          Odysseas Sofikitis\inst{2,3} \and
          Christos Stefo\inst{1}}
  \authorrunning{Z. Avarikioti et al.}
  \institute{TU Wien \and
             Common Prefix \and
             National Technical University of Athens}
\fi

\begin{document}
\graphicspath{{./figures/}}

\maketitle

\begin{abstract}
State machine replication protocols are usually proven secure in the
honest-Byzantine setting, which requires an honest supermajority of validators.
Deployed systems offer no such guarantee: validators participate for profit and
may deviate from the protocol whenever doing so increases their gains. In our
work, every validator is rational or Byzantine, and none is honest. Rational
validators may collude with each other and with the Byzantine ones, who may
also bribe them to deviate. Known impossibility results show that, without
honest validators, clients that decide whether to accept a transaction based on
a non-interactive certificate and without further communication cannot be kept
safe. To bypass these results, we let clients relay the certificates they
receive to full nodes and wait before confirming transactions. We also assume
that one full node, whose identity is unknown, is honest and always online.

Under this assumption, we build a finalization layer over any certifiable
quorum-based protocol in a synchronous network. Clients run an existing freezing
gadget and fetch past certificates from the honest full node, which makes the
protocol \emph{always-safe}: no two honest clients confirm conflicting ledgers, for any
strategy of the validators. We incentivize participation through rewards, which
the adversary may use to bribe rational validators into censoring transactions.
In this setting, we formalize censorship and characterize it tightly: paying
only with his rewards, the adversary censors less than a $\bw$ fraction of the
available fees at each height, where $\bw$ is the fraction of the quorum he
signs, and every coin of censorship beyond that comes from his own funds.
Finally, we extend the construction to dynamic validator committees. There, a
client that sleeps no longer than stake stays slashable keeps the guarantees of
the static case, and, when a committee may choose its successor, no safe
protocol that is live for always-on clients guarantees liveness to a client
that sleeps longer.

\end{abstract}

\section{Introduction}\label{sec.introduction}

State machine replication (SMR) lets a set of validators agree on a single ordered log of transactions, and it is the core of proof-of-stake blockchains that today secure hundreds of billions of dollars. Quorum-based SMR protocols such as PBFT~\cite{pbft}, Tendermint~\cite{tendermint}, and HotStuff~\cite{hotstuff} are typically analyzed in the \emph{honest-Byzantine} setting: every validator either follows the protocol or deviates arbitrarily. These protocols remain safe and live as long as fewer than a third of the validators are Byzantine, so their security rests on an honest supermajority. Clients rely on the same assumption when they accept a transaction from a quorum certificate: the quorum is guaranteed to contain honest validators that have followed the protocol.

This setting may not describe how validators behave in deployed proof-of-stake systems. Validators participate for profit: they follow the protocol when doing so pays, and may deviate when they can earn more by doing so. An adversary may offer them a bribe to deviate, or validators may collude by pooling their stake and profiting from a double-spend. Thus, the honest supermajority may fail not because validators are intrinsically malicious, but because deviating has become more profitable than following the protocol. An analysis in the honest-Byzantine setting provides no guarantees in this case. This raises the question we study: what happens to the security of a quorum-based SMR protocol when every validator is rational or Byzantine, and no validator is assumed to be honest?

\subsection{Related Work}\label{sec.related}
\noindent\textbf{Rational and Byzantine validators.}
BAR~\cite{bar} introduced a model with Byzantine, altruistic, and rational participants, and a line of work since then has studied consensus with rational validators~\cite{rational-vs-byzantine,rational-committee,tenderstake,fairledger}. These works either assume that rational validators act independently or rule out collusion between rational and Byzantine validators against clients. Abraham et al.~\cite{adgh06,adh08} introduced solution concepts that allow coalitions of rational and Byzantine players, and TRAP~\cite{trap} used these concepts to solve consensus on a single value by baiting rational players into exposing a coalition. In TRAP, the single-shot setting bounds the amount that can be double-spent against clients. Multi-shot state machine replication has no such bound, as shown by CoBRA~\cite{cobra}. CoBRA instead considers quorum-based SMR with silent clients and shows that, without honest validators, safety is impossible even in a synchronous network. Here, a \emph{silent client} is one that decides whether to confirm a transaction based only on the certificates it receives, without communicating with other parties. Such a client cannot tell whether a certificate it sees belongs to a history that conflicts with one another client has accepted.

\noindent\textbf{Safety beyond the honest majority.} Another line of work keeps clients safe when Byzantine validators exceed the classic thresholds, by changing what clients do. Sridhar et al.~\cite{adversary-majority} show that the achievable resilience depends on the client model, whether clients are sleepy or always on, and silent or communicating, and characterize it for each combination. In flexible BFT~\cite{flexbft,flexopt}, clients trade liveness for safety by requiring larger quorums. Better Safe than Sorry~\cite{bettersafe} introduces the freezing gadget we build on: a client that sees two conflicting histories confirms neither, which keeps clients safe under an adversarial majority, and a recovery procedure restores progress once an honest majority returns. Sridhar et al.\ prove the gadget safe against any number of adversarial validators when past messages reach a client that wakes~\cite[Theorem 4]{adversary-majority}, and note that querying online parties provides this. Our safety theorem instantiates theirs with one honest full node. In both works liveness is lost while the adversary holds the majority, and in a purely Byzantine model this cannot be avoided, since Byzantine validators can simply stop voting.

\noindent\textbf{Accountability and recovery.} Accountable protocols~\cite{polygraph,forensics,abc,accgadget} identify the validators responsible for a safety violation so that their stake can be slashed, and recoverable protocols~\cite{zlb,gong,blr} restore a consistent state after a violation. Both act after the violation. Punishment deters a rational validator only when the gain from an attack is smaller than the stake it risks, and in multi-shot SMR the gain from a double-spend has no natural bound.

\noindent\textbf{Censorship.} The cost of censorship by an outside briber has been studied for a single proposer~\cite{fox-pai-resnick}, and concurrent proposers raise it~\cite{price-of-censorship,tfm-multiproposer,focil}; colluding proposers are left open~\cite{price-of-censorship}. CoBRA~\cite{cobra} observes that Byzantine validators can use their participation rewards to bribe rational ones. We quantify this effect: the share of the fees that the adversary earns by signing is exactly what he can spend on censorship at every height.

\noindent\textbf{Changing committees.} A committee that has withdrawn its stake can sign a competing history at no cost. Snow White~\cite[Theorem 2]{snowwhite-full}, Tas et al.~\cite[Theorem 1]{tas-babylon} and Lewis-Pye and Roughgarden~\cite[Theorem 10.1]{lpr-permissionless} show that a client that joins late cannot then identify the correct history without further trust. Deployed systems therefore require a recent checkpoint~\cite{winkle} and bound the trusting period of light clients by the unbonding period~\cite{tendermint-lc}. We carry this bound over to rational validators and an honest full node, and show that it is exact.

\subsection{Our Work}
The lines of work above leave a gap. Work on rational and Byzantine validators models profit-driven validators that may collude, but it retains honest validators; CoBRA shows that, with silent clients, such an assumption is necessary. Work on accountability, recovery, and safety beyond the Byzantine threshold treats every validator that is not honest as Byzantine, and therefore gives up liveness while such validators hold a majority. Our work lies in the middle: we keep the Byzantine validators within the resilience of the underlying protocol, while every other validator is rational and may collude with the adversary; none is honest.

\noindent\textbf{Assumptions.} Within these limits, we ask what minimal assumptions at the client level suffice to solve SMR. We make two assumptions, each of which removes a known impossibility. First, the network is synchronous with a known bound $\Delta$, and clients communicate: a client relays every certificate it sees to the full nodes and waits before confirming. CoBRA shows that, without honest validators, no protocol is secure under partial synchrony~\cite[Theorem 3.1]{cobra}, and that even under synchrony, clients cannot remain silent when the number of rational and Byzantine validators exceeds the quorum threshold~\cite[Theorem 3.5]{cobra}. Synchrony escapes the first result, and a client that relays certificates and waits a known multiple of $\Delta$ falls outside the second. Second, we assume the existence of full nodes that store the certificates they receive and provide them to clients on request. We assume \emph{existential honesty}: at least one full node is honest and has been online since the start of the execution, every client can reach it, and no party knows which one it is. Sleepy clients cannot be kept both safe and live once half of the validators are adversarial~\cite[Theorem 3]{adversary-majority}, while always-on clients can~\cite[Corollary 3]{adversary-majority}; the honest full node is an always-on party that sleepy clients consult. Full nodes hold no stake and cast no votes, and an honest validator that stores and serves certificates could play the role of this full node, so existential honesty is weaker than assuming one such honest validator.

\noindent\textbf{Contributions.}
We formulate an SMR model in which every validator is rational or Byzantine:
rational validators may collude with the adversary and accept his bribes,
clients are sleepy, and the only honest party is an always-on full node
whose identity is unknown. The first two contributions concern a static
committee, where clients may sleep arbitrarily long and join at any time; the
third lets the committee change, which forces a bound on how long clients may
sleep.
\begin{itemize}
\item \emph{A tight characterization of censorship.} We define the
adversary's \emph{bandwidth} as the fraction of a quorum certificate that
he signs. Paying only with the rewards of a height, he censors less than his
bandwidth times the fee value available, and every further unit of censored
fees costs him a unit of his own budget, tightly (\Cref{thm.censorship}).
Conflicting certificates cost him more than a validator's stake minus one
block reward (\Cref{lem.equivocation-cost}), so a smaller budget never freezes
a client, and a freeze is permanent. Transactions above the resulting fee
threshold are included within a bounded number of certified heights
(\Cref{cor.inclusion,thm.liveness}).
\item \emph{An always-safe finalization layer.} We build a finalization layer
that runs on top of any certifiable quorum-based protocol under synchrony
and changes only how ledgers are finalized (\Cref{sec.construction}).
Clients use the freezing gadget of \cite{bettersafe}: they relay every
certificate to the full nodes, wait $2\Delta$, and confirm a ledger only if
no conflicting ledger reaches them. Full nodes serve past certificates to
clients that were offline. Adapting \cite[Theorem 4]{adversary-majority}, no
two honest clients confirm conflicting ledgers, whatever the validators do
(\Cref{thm.safety}).
\item \emph{An extension to dynamic committees.} We extend the construction
to committees that change every epoch (\Cref{sec.dynamic}). Safety holds for
every client, whenever it joins. Liveness holds for a client that sleeps no
longer than the number of epochs for which stake stays slashable
(\Cref{thm.dynamic-liveness}). Adapting the classical lower bound for
long-range attacks, we show that this threshold is exact when a committee may
choose its successor, even with an honest full node and against any positive
budget (\Cref{thm.freshness-necessary}).
\end{itemize}

\noindent\textbf{Organization.}
\Cref{sec.model} fixes the model and the definitions. \Cref{sec.construction}
gives the construction for the static validator set and \Cref{sec.analysis}
analyzes it. \Cref{sec.dynamic} extends the construction to changing
committees, and \Cref{sec.discussion} discusses its limits. The proofs are in
the appendix.

\section{Model}\label{sec.model}

\subsection{System Model}

We consider three types of actors: \emph{validators}, \emph{clients}, and
\emph{full nodes}. We define the three types of actors precisely below.

\noindent\textbf{Validators.} Validators run the SMR protocol \prot to maintain
an ordered ledger of transactions. In \Cref{sec.construction}, we consider a
\emph{static} setting, in which there is a fixed set $\val$ of $n$ validators
known to all parties. In \Cref{sec.dynamic}, we consider a \emph{dynamic}
setting, in which the execution is divided into epochs, and we denote the
validator committee of epoch $e$ by $\comm_e$. In both settings, each validator
holds a signing key pair, and every party knows the public key of every
validator. Validators are always awake. Each validator locks $\stake$ coins as stake. We model an entity that
locks $\alpha \cdot \stake$ coins as controlling $\alpha$ validators.

\noindent\textbf{Full nodes.} Full nodes store certificates, relay them, and
send them to any party that requests them. They hold no stake, cast no votes,
and earn no rewards. They are not registered as part of the system and
therefore require no PKI; their number is unbounded and unknown.

\noindent\textbf{Clients.} Clients use the ledger by holding coins, submitting
transactions, and deciding whether to accept transactions, but they do not
participate in the execution of \prot. An honest client submits a transaction
by sending it to every validator. Clients are \emph{sleepy}: they may remain
offline for extended periods and receive no messages while asleep. Clients
communicate with validators and full nodes, but not with each other. We assume
that at least one full node, which we denote \fullnode, is honest and has been
awake since the beginning of the execution, and that it and every client can
reach each other. No party knows which full node is \fullnode. We call this assumption
\emph{existential honesty}.

\noindent\textbf{Time and network.} Time proceeds in rounds, and parties have
synchronized clocks. The network is synchronous with a known delay $\Delta$: a
message sent at round $r$ to a party $p$ reaches $p$ by round $r+\Delta$,
provided $p$ is awake from round $r$ to round $r + \Delta$. A sleeping party
receives nothing, and messages it missed are not delivered when it wakes. A
party \emph{multicasts} a message by sending it to every party it communicates
with.

\noindent\textbf{Assumptions.}
We assume unforgeable digital signatures and collision resistant hash
functions. Moreover, we assume that a genesis block $G$, available to every party,
specifies the initial validators, their public keys, and their stake.

\noindent\textbf{Adversarial model.}
In the static setting, $|\val|=3f+1$; in the dynamic setting, every committee
satisfies $|\comm_e|=3f+1$. In either case, $f$ validators are Byzantine and the
remaining $k=2f+1$ validators are rational. Byzantine validators may behave
arbitrarily, while rational validators maximize their \emph{monetary profit},
which we define precisely in \Cref{sec.utilities}. Rational validators may
collude with one another and with the adversary, accept bribes from the
adversary, and deviate from the protocol whenever doing so increases their
profit.

The adversary \A controls all Byzantine validators and may corrupt any number
of clients and of full nodes other than \fullnode. He controls message
delivery subject to the synchronous bound $\Delta$ and is \emph{rushing}: in
each round, he observes the messages sent by other parties before choosing his
own actions. The adversary starts with a budget of \budget\ coins that are not
staked. He also receives the rewards of his validators, and we write $g_h$ for
the reward he receives from the block of height $h$. Given enough heights,
saved rewards pay for any finite attack. We therefore charge each side payment
to the height of the action it buys, whenever it is paid, and measure the
payments charged to height $h$ against $g_h$. Any excess is drawn from
\budget. We write
$\budget_h \geq 0$ for the amount drawn at height $h$, so that
$\sum_h \budget_h \leq \budget$. Fees that the adversary pays on his own
transactions count as side payments.

\begin{definition}[Repeatable attack]\label{def.repeatable}
    An adversarial strategy is \emph{repeatable} at height $h$ if
    $\budget_h = 0$, that is, if the side payments it makes at height $h$ are
    at most $g_h$.
\end{definition}

A repeatable strategy can run at every height. A strategy that is not
repeatable draws on \budget, which rewards do not replenish.

We make one assumption on how certificates are formed, which we call
\emph{adversarial participation}. The adversary chooses how many of his
validators, $f^{*} \leq f$, sign the certificate of a block, and their
signatures are included. A rational signer of a block $b$ compares
signing $b$ with signing $b$ extended by the transactions it holds. The $k = q$
rational validators can certify the second block without the adversary, and it
pays each signer $1/q$ of its fees. A rational validator recruited to replace a
signer that refuses makes the same comparison.

\noindent\textbf{Notation.} Throughout, we use the following terminology for
ledgers and blocks. A block $b=(h,H(b'),T)$ has height $h$, the hash $H(b')$ of
its parent $b'$, and an ordered sequence $T$ of transactions. Two blocks
\emph{conflict} if neither extends the other. A ledger \ledger\ is a sequence of
blocks. We write $\ledger^c_r$ for the ledger output by client $c$ by round
$r$. We write $\ledger\preceq \ledger'$ if \ledger\ is a prefix of $\ledger'$.
Two ledgers are \emph{consistent}, written $\ledger\sim \ledger'$, if one is a
prefix of the other, and \emph{conflicting}, written $\ledger\not\sim
\ledger'$, otherwise. Two conflicting ledgers contain conflicting blocks at the
first position where they differ.

\noindent\textbf{Certifiable protocols.} Our transformation applies to any
protocol \prot for which a party that did not monitor the execution can verify
the protocol's output using a non-interactive certificate. We refer to such
protocols as \emph{certifiable protocols}, as defined in~\cite{bettersafe}.

\begin{definition}[Certifiable protocol~\cite{bettersafe}]
\label{def.certifiable}
An SMR protocol \prot is \emph{certifiable} if it comes with a witness
producer \Wprod and a deterministic non-interactive witness consumer
\Wcons: a party invoking \Wprod at round $r$ obtains a witness $w$
with $\Wcons(w)=\ledger$, the ledger \prot has output at that party by round
$r$, and $\Wcons(w)=\bot$ for every $w$ that is not a valid witness.
\end{definition}

\noindent\textbf{Quorum-based protocols.} We focus on protocols in which a
block $b$ is finalized after a set of $q = 2f+1$ signatures on $b$ by distinct
members of the committee, such as PBFT~\cite{pbft},
HotStuff~\cite{hotstuff} and Tendermint~\cite{tendermint}. Validators that
follow \prot\ certify at most one block per height. We call the certificate that finalizes
$b$ a \emph{valid quorum certificate}, denote it by $\qc_b$, and write
$\signers(\qc_b)$ for the members whose signature is on $\qc_b$. A witness for a
ledger $\ledger=(b_0,\ldots,b_h)$ consists of the ledger together with a quorum
certificate for each block:
$w=\big(G,(b_0,\qc_{b_0}),\ldots,(b_h,\qc_{b_h})\big)$. The consumer \Wcons
checks every certificate and every parent hash, and executes the ledger from
$G$. It returns $\bot$ if a check fails or a transaction is invalid, since no
quorum is guaranteed to contain a validator that checked the block.

\noindent\textbf{Equivocations and accountability.} We assume that \prot is
accountable: the quorum certificates $\qc_b$ and $\qc_{b'}$ for any two
conflicting blocks $b$ and $b'$ prove that every validator in their
intersection $v \in \signers(\qc_b) \cap \signers(\qc_{b'})$ has deviated from
the protocol specification. The pair $(\qc_b, \qc_{b'})$ is the
\emph{evidence} of the equivocation. A validator that equivocates loses its
stake \stake. Most quorum-based protocols are accountable or become so with
small changes~\cite{forensics,polygraph}, and any consensus protocol can be
made accountable by a generic transformation~\cite{abc}.

\noindent\textbf{Transactions and fees.} We assume that every transaction \tx\
submitted by clients carries a fee $\tx.\fee\ge 0$. We write $\mempool_r^p$ for
the set of transactions pending in the view of validator $p$ during the round
$r$, and $\reward(X)=\sum_{\tx\in X}\tx.\fee$ for the total fee value of a set
$X$ of transactions, such as a block $b$ or the set $\mempool_r^p$. A block may
contain any number of transactions, and including a transaction in a block
costs its proposer nothing. We write $\maxfee$ for an upper bound on the fee
value of a block and assume $\stake > \maxfee$.

\subsection{Ledger Properties}
\label{sec.ledger-properties}

We consider the standard state machine replication (SMR) problem, but state its
properties from the perspective of clients. In the usual formulation, safety
and liveness are defined with respect to the ledgers of correct validators,
together with a requirement that clients can learn those ledgers. In our
model, however, no validator is assumed to be correct: validators may be
Byzantine or rational and may deviate from the protocol. We therefore state
the ledger properties directly over the ledgers output by honest clients,
following the perspective adopted in~\cite{bettersafe,adversary-majority}. We
first define safety and liveness and then introduce a weaker liveness notion
that tolerates transaction censorship.

\begin{definition}[Safety]
\label{def.safety}
A protocol satisfies \emph{safety} if, for all honest clients $c_1,c_2$ and
all rounds $r_1,r_2$, the ledgers of $c_1$, $c_2$ are consistent, i.e.,
$\ledger^{c_1}_{r_1}\sim \ledger^{c_2}_{r_2}$.
\end{definition}

\begin{definition}[Liveness]
\label{def.liveness}
A protocol satisfies \emph{liveness} if, for every transaction \tx\ submitted
by an honest client at round $r$, there exists a round $r'>r$ such that every
honest client that outputs a ledger at a round $r''\geq r'$ outputs a ledger
containing \tx.
\end{definition}

\noindent\textbf{Censorship.} The liveness property above is too strong for our
setting, where the adversary may use his budget to bribe rational validators
and thereby prevent some transactions from being included in certified blocks.
We therefore define \emph{$\gamma$-censorship resistance}, which ensures that
the value of transactions that can be censored at any height is bounded by a
$\gamma$-fraction of the transaction value available to the rational signers,
in addition to the adversary's budget. We measure the adversary's share of a
certificate as follows.

\begin{definition}[Adversarial bandwidth]\label{def.bandwidth}
    Let $\qc_b$ be a quorum certificate and let $f^{*}$ be the number of
    Byzantine validators among its signers. The \emph{adversarial bandwidth} of
    $\qc_b$ is $\bw = f^{*}/q$.
\end{definition}

\begin{definition}[Censored value]\label{def.censored}
Let $b$ be the block certified at height $h$, proposed at round $r$, and let
$\mathcal{R}$ be the set of rational signers of $\qc_b$. Let
$\mempool_h = b \cup \bigcap_{p \in \mathcal{R}} \mempool^p_r$ be the transactions of $b$
together with the transactions held by every rational signer. The
\emph{censored value} at height $h$ is
$\cens_h = \reward(\mempool_h) - \reward(b)$.
\end{definition}

\begin{definition}[$\gamma$-censorship resistance]\label{def.censorship-resistance}
Fix $\gamma \in [0,1)$. A protocol is \emph{$\gamma$-censorship resistant}
if $\cens_h \leq \gamma \cdot \reward(\mempool_h) + \budget_h$ at every
height $h$ at which a block is certified.
\end{definition}

\subsection{Utilities}
\label{sec.utilities}

\noindent\textbf{The game.}
An execution of \prot induces a game whose players are the validators. A
strategy maps a validator's local view to its actions. Byzantine validators
may also make side payments to other validators, subject to the adversary's
budget.

\noindent\textbf{Utility.} For any execution and any interval $I$ of that
execution, the utility of a rational validator $p$ is
$
u_p^I=\mathrm{rew}_p^I+\mathrm{bribe}_p^I+\mathrm{gain}_p^I-\mathrm{pen}_p^I$,
where $\mathrm{rew}_p^I$ is the protocol reward received by $p$ during $I$,
$\mathrm{bribe}_p^I$ is the side payment received by $p$ from the adversary,
$\mathrm{gain}_p^I$ is its share of any double-spent value (if any such attack
is successful), and $\mathrm{pen}_p^I$ is the stake lost to slashing. A
validator receives the reward of a block only once an honest client confirms
the block. A side payment is made before the action it buys, and a rational
validator that accepts it performs that action; a promise of a future payment
does not enter the utility. A rational validator deviates only when doing so
strictly increases its utility and follows the protocol when indifferent. We
assume that it propagates every certificate it holds unless withholding one
strictly increases its utility.

\section{Construction}\label{sec.construction}

The construction is a finalization layer over \prot. Validators run \prot
unchanged and send every certificate they hold to the full nodes. Full nodes
run \Cref{alg.fullnode}, and clients run the freezing gadget of
\cite{bettersafe} (\Cref{alg.client}).

\noindent\textbf{Full nodes.}\label{sec.construction.fullnodes} A full node
keeps a store of certificates indexed by height. On receiving a message it
stores the certificates it did not hold (\Cref{line.fstore}) and relays them
(\Cref{line.relay}), before any other action. On receiving a request it sends
its store (\Cref{line.serve}). Full nodes confirm nothing.

\noindent\textbf{Clients.}\label{sec.construction.clients} A client that wakes
multicasts a request (\Cref{line.crequest}). On every message it stores the
certificates and relays the message (\Cref{line.cstore,line.cgossip}). Every
chain of certified blocks in $R$ that starts at genesis is a witness. For every
witness with a ledger not yet in \seen, the client adds the ledger to \seen\
(\Cref{line.cseen}) and schedules \fn{confirm} for round $t + 2\Delta$, where
$t$ is the current round (\Cref{line.cschedule}). \fn{confirm} outputs the
ledger if \seen\ holds no conflicting ledger and the ledger extends \out\
(\Cref{line.cconflict,line.cextend}). A client that has seen two conflicting
ledgers confirms neither, nor any ledger that extends either. We say it
\emph{freezes}. A frozen client keeps relaying. Within a round, a client
processes messages before scheduled confirmations. A client that falls asleep
keeps \out\ and discards $R$, \seen\ and its pending confirmations.

The client waits $2\Delta$, one $\Delta$ more than in \cite{bettersafe},
because clients reach each other only through full nodes: a relayed
certificate reaches \fullnode within $\Delta$, and its relay reaches every
awake client within another $\Delta$. The same $2\Delta$ covers the request
round trip of a client that has just woken. Here \fullnode delivers the past
messages that the network of \cite{bettersafe,adversary-majority} delivers.

\noindent\textbf{Validity.} A quorum may certify a block that moves coins their
owner never had, which creates no conflict for the freezing gadget to detect.
The consumer \Wcons therefore executes the whole ledger of a witness and
rejects invalid transitions (\Cref{sec.model}).

\noindent\textbf{Rewards.}\label{sec.construction.fees} A block pays the fees
it carries, split equally among the $q$ signers of its certificate, once an
honest client confirms it. A proposer-only reward would let the adversary stall
a block by bribing one signer for any positive amount. A signer is paid only if
the certificate of its block reaches the clients, so a rational signer sends it
to the full nodes.

\noindent\textbf{Intuition.} Safety needs no incentive. When a client confirms a
ledger, \fullnode has already shown its certificates to every awake client, and
serves them to every client that wakes later, so a client that would confirm a
conflicting ledger afterwards freezes instead. Incentives
keep clients from freezing and blocks from being censored. Two conflicting
certificates share $f+1$ signers, so at least one rational validator signs
both and loses its stake \stake. The adversary must pay it more than his
reward from one height, which only \budget\ can cover. Omitting a transaction from
a block lowers the reward of every signer. The adversary must compensate the
rational signers, and his own share of the fees is what he can spend on
side payments. A repeatable strategy therefore censors less than a \bw\
fraction of the available fees (\Cref{sec.analysis}).

\begin{algorithm}[t]
\caption{The relay run by every full node. A full node stores every certificate
it receives, relays it at once, and serves its store on
request.}\label{alg.fullnode}
\small
\begin{algorithmic}[1]
    \State $\quorumstore \gets \emptyset$                               \Comment{Certificates received, indexed by height}

    \Statex
    \On{\receive\ a message $m$}\label{line.frecv}
        \State $Q \gets$ the valid certificates carried by $m$ that are not in \quorumstore
        \If{$Q = \emptyset$}\label{line.fdedup}                          \Comment{Nothing new}
            \return
        \EndIf
        \State $\quorumstore \gets \quorumstore \cup Q$\label{line.fstore}
        \State $\multicast(Q)$\label{line.relay}                          \Comment{Relay before anything else}
    \EndOn

    \Statex
    \On{\receive\ \request\ from party $P$}\label{line.serve}
        \State send \quorumstore\ to $P$
    \EndOn
\end{algorithmic}
\end{algorithm}

\begin{algorithm}[t]
\caption{The freezing gadget of \cite{bettersafe}, run by a client. Every
message is relayed on arrival. The ledger of a witness is confirmed $2\Delta$
after the witness is complete, if no conflicting ledger has been
seen.}\label{alg.client}
\small
\begin{algorithmic}[1]
    \State $R \gets \emptyset$                                          \Comment{Certificates received}
    \State $\seen \gets \emptyset$                                      \Comment{Ledgers seen}
    \State $\out \gets \emptyset$                                       \Comment{Confirmed ledger, the output}

    \Statex
    \On{waking}\label{line.cwake}
        \State $\multicast(\request)$\label{line.crequest}               \Comment{Ask the network for their store}
    \EndOn

    \Statex
    \On{\receive\ a message $m$ at round $t$}\label{line.crecv}
        \State $R \gets R \cup \{\text{valid certificates carried by } m\}$\label{line.cstore}
        \State $\multicast(m)$\label{line.cgossip}                       \Comment{Relay before anything else}
        \ForAll{witnesses $w$ formed by certificates in $R$ with $\Wcons(w) \notin \seen$}\label{line.cwitness}
            \State $\seen \gets \seen \cup \{\Wcons(w)\}$\label{line.cseen}
            \State $\schedule(\fn{confirm}, t + 2\Delta, \Wcons(w))$\label{line.cschedule}
        \EndFor
    \EndOn

    \Statex
    \Function{\fn{confirm}}{\ledger}
        \If{$\exists\, \ledger' \in \seen \colon \ledger \not\sim \ledger'$}\label{line.cconflict} \Comment{Conflict seen: do not confirm}
            \return
        \EndIf
        \If{$\out \preceq \ledger$}\label{line.cextend}
            \State $\out \gets \ledger$\label{line.coutput}              \Comment{Output}
        \EndIf
    \EndFunction
\end{algorithmic}
\end{algorithm}

\section{Analysis}\label{sec.analysis}

We prove the results of this section in \Cref{app.static}.
\Cref{lem.equivocation-cost} bounds the cost of a freeze, \Cref{thm.censorship}
the cost of censorship, and \Cref{thm.liveness} combines them.

\subsection{Safety}\label{sec.analysis.safety}

\begin{theorem}[Safety]\label{thm.safety}
    The construction of \Cref{sec.construction} satisfies safety
    (\Cref{def.safety}) for every strategy of the validators.
\end{theorem}

The proof follows~\cite[Theorem 4]{adversary-majority}. The new step is
\Cref{lem.delivery}: a certificate that an honest client relays at round $t$
reaches, through \fullnode, every honest client that confirms at a round
$s \geq t + 2\Delta$, even one asleep at round $t$. In particular, the term
$\mathrm{gain}_p^I$ of every rational validator is zero.

\subsection{Liveness}\label{sec.analysis.liveness}

\begin{lemma}[Cost of equivocation]\label{lem.equivocation-cost}
    In the static setting, if two conflicting blocks are certified, then at
    least $f+1$ validators equivocate and at least one of them is rational. In
    both settings, under client freshness in the dynamic one, for every
    rational validator that loses its stake for an equivocation, the adversary
    pays it
    more than $\stake - \maxfee/q$ before it signs and draws more than
    $\stake - \maxfee$ from \budget, so the attack is never repeatable. In
    particular, if $\budget \leq \stake - \maxfee$, then no two conflicting
    blocks are certified and no honest client freezes.
\end{lemma}

Two quorums share $f+1$ members, at least one rational, which loses \stake\ and
gains at most one block reward.

\begin{theorem}[Censorship]\label{thm.censorship}
    Assume adversarial participation. At every height $h$ at which a block is
    certified with bandwidth \bw\ and $\cens_h > 0$,
    \[
        \cens_h < \bw \cdot \reward(\mempool_h) + \budget_h \,.
    \]
    The bound is tight. Suppose the rational signers hold the same
    transactions. For every set of transactions in $\mempool_h$ of total fee
    $c > 0$ and every $\varepsilon > 0$, the adversary can censor exactly this
    set at height $h$ with
    $\budget_h \leq \max(0,\, c - \bw\reward(\mempool_h) + \varepsilon)$ once the
    block is confirmed.
\end{theorem}

Every rational signer of a censoring block loses its share of the censored
fees and must be paid more than that. The adversary pays with his own share
\bw\ of the fees, so, when the rational signers hold the same transactions,
censoring fee value $c$ at a height is repeatable if and only if
$c < \bw\reward(\mempool_h)$. Since $\bw \leq f/q$, the construction
is $\frac{f}{q}$-censorship resistant (\Cref{def.censorship-resistance}), and
censoring all of $\mempool_h$ is never repeatable when
$\reward(\mempool_h) > 0$. Without side payments, no rational signer signs a
block that omits a transaction with a positive fee that it holds.

\begin{corollary}[Price of inclusion]\label{cor.inclusion}
    Let a block $b$ be certified at height $h$ with bandwidth \bw, and let
    $\tx \in \mempool_h$ with $\tx.\fee > 0$. Then $\tx \in b$ whenever
    \[
        \tx.\fee \geq \frac{\bw}{1-\bw}\reward(\mempool_h \setminus \{\tx\})
            + \frac{\budget_h}{1-\bw} \,.
    \]
    Since $\bw \leq \frac{f}{2f+1}$, it suffices that
    $\tx.\fee \geq \frac{f}{f+1}\reward(\mempool_h \setminus \{\tx\})
        + \frac{2f+1}{f+1}\budget_h$.
\end{corollary}

\begin{theorem}[Liveness]\label{thm.liveness}
    Assume adversarial participation and $\budget \leq \stake - \maxfee$. Then
    no honest client freezes, and the following hold.
    \begin{enumerate}
        \item If the certificates of a valid ledger \ledger\ have reached \fullnode
        by round $s$, then every honest client that is awake throughout
        $[s', s' + 4\Delta]$ for some $s' \geq s$ outputs, by round
        $s' + 4\Delta$, a ledger that extends \ledger.
        \item Let \tx\ be submitted by an honest client at round $r$, and let
        $\delta > 0$ be such that
        $\tx.\fee \geq \frac{f}{f+1}\reward(\mempool_h \setminus \{\tx\})
        + \delta$ at every height $h$ whose block is proposed after round
        $r + \Delta$. Then at most $2\budget/\delta$ such heights certify a
        block without \tx\ before \tx\ is included.
    \end{enumerate}
\end{theorem}

\begin{remark}\label{rem.liveness-limits}
    \Cref{thm.liveness} is weaker than liveness (\Cref{def.liveness}). By
    \Cref{thm.censorship}, a transaction whose fee stays below
    $\frac{\bw}{1-\bw}\reward(\mempool_h \setminus \{\tx\})$ at every height
    can be excluded at every height by a repeatable strategy when the rational
    signers hold the same transactions. We claim safety
    in full and liveness only for transactions that clear this threshold,
    which is $\frac{f}{f+1}$ of the other pending fees when $\bw = f/q$. A stall,
    or a certificate withheld from every full node, pays the adversary nothing
    and needs a paid rational validator, so neither is repeatable. We do not
    bound the delay they cause.
\end{remark}

\section{Changing Committees}\label{sec.dynamic}

We now let the validator set change. Each committee signs a handover
certificate that fixes the last block of its epoch, and each client validates
forward from an anchor taken from the last ledger it confirmed. Safety needs nothing
more (\Cref{thm.dynamic-safety}). Liveness holds when stake stays slashable for
$\bwin$ epochs and no client sleeps longer (\Cref{thm.dynamic-liveness}), and
the second bound is necessary (\Cref{thm.freshness-necessary}).

\subsection{Setting and Construction}\label{sec.dynamic.construction}

\noindent\textbf{Epochs and committees.}
Fix an epoch length $E \in \mathbb{N}$. Epoch $e$ consists of the rounds $[eE, (e+1)E)$. A block
carries the round at which it is proposed and belongs to the epoch of that
round, and the rounds of the blocks of a ledger increase with height. We write
$h_e$ for the last height of epoch $e$, that is, the height of the last block
whose epoch is at most $e$, and $\sigma_e$ for the state after the block of
height $h_e$. The committee of epoch $e$ is
$\comm_e = \committee(\sigma_{e-1})$, where $\committee$ is a fixed public
function, and $\comm_0$ is the genesis committee. We abstract how transactions
determine the committee. During epoch $e$ the members of $\comm_e$ run \prot,
and a quorum certificate for a block of epoch $e$ carries $q$ signatures of
distinct members of $\comm_e$. How the members of $\comm_e$ obtain the ledger
at the start of the epoch is part of \prot. Full nodes run the static
construction and store every well-formed certificate, since they track no
committees.

\noindent\textbf{Handover certificates.}
The transition from epoch $e-1$ to epoch $e$ is certified by the outgoing
committee. Every member of $\comm_{e-1}$ signs the pair $(e, H(b))$, where $b$
is the block of height $h_{e-1}$ (\Cref{line.hosign}). A quorum of $q$ such
signatures forms the handover certificate $\handover_e$
(\Cref{line.hoquorum}), which the first block above height $h_{e-1}$ includes
(\Cref{line.hoinclude}). A certificate for a block of epoch $e-1$ that extends
$b$, together with $\handover_e$, is evidence of an equivocation by every
validator that signed both, as is a pair of certificates for conflicting
blocks.

\noindent\textbf{Slashing.} Fix $\bwin \in \mathbb{N}$. The stake of every
validator in $\comm_e$ remains locked, and slashable on evidence of an
equivocation on the blocks of epoch $e$ or on $\handover_{e+1}$, until
$5\Delta$ rounds after the end of epoch $e + \bwin$. We call this period the
\emph{bonding window} of $\comm_e$. A committee that receives such evidence
within the period includes it in a block of its epoch, and its members share
the slashed stake equally. The bound on Byzantine members of \Cref{sec.model}
holds for the committees that a ledger designates when the quorum certificate
of every block of the ledger was formed within the bonding window of its
signers; other ledgers may designate any keys.

\noindent\textbf{Anchors.}
A client checks the certificates of epoch $e$ against $\comm_e$, which it
learns only after validating epoch $e-1$, and so on recursively from the
genesis. It therefore needs a point in the history from which to validate
forward every time it wakes up. To this end, the client keeps a commitment to
the last epoch it validated before sleeping. We call this point the client's
\emph{anchor}. The anchor of a client is a triple $(e_0, \comm_{e_0}, \eta_0)$:
an epoch $e_0$, its committee $\comm_{e_0}$, and the hash $\eta_0$ of the block
of height $h_{e_0-1}$. The client takes its anchor from the last ledger $a$ it
confirmed, and $e_0$ is the epoch of the last block of $a$. A client that has
never confirmed a ledger holds the genesis anchor $(0, \comm_0, \bot)$. The
hash $\eta_0$ fixes the block of height $h_{e_0-1}$ and, through the parent
hashes, every block below it. The committee $\comm_{e_0}$ alone would not fix
them, since two different ledgers may designate the same committee.

\noindent\textbf{Clients.}
The client keeps running finalization of \Cref{alg.client} unchanged except for
the witness consumer $\Wcons$ of the certifiable protocol. \Cref{alg.handover}
gives the witness consumer $\bootstrap$, which verifies the committee changes
across epochs. The client initializes $\out$ to $a$ and keeps its anchor, the
state after $a$, and \out\ across sleeps. A witness carries the blocks, certificates
and handover certificates from height $h_{e_0-1}$ onward.  $\bootstrap$ checks
that the witness extends the block that $\eta_0$ fixes, with increasing rounds
and no block of an epoch below $e_0$ above it (\Cref{line.boanchor}).  It then
validates one epoch at a time from $e_0$: every block of epoch $e'$ is valid and
carries a quorum certificate of $\comm_{e'}$ (\Cref{line.boverify}), the next
committee is derived from the state (\Cref{line.bonext}), and $\handover_{e'+1}$
is a quorum of $\comm_{e'}$ (\Cref{line.bohandover}). The ledger of a witness is
the prefix of $a$ that ends at height $h_{e_0-1}$, followed by the blocks that
$\bootstrap$ returns above it.

\noindent\textbf{Assumptions.}\label{sec.dynamic.assumptions}
The liveness results of this section rest on the bonding window and on one
assumption on clients, which we call \emph{client freshness}. An honest client
that wakes during epoch $e$ holds an anchor of epoch $e_0 \geq e - \bwin$. While
it stays awake, its anchor must stay equally recent, except during the first
$4\Delta$ rounds after it wakes, which it needs to confirm a new ledger: at
every round $r$ at least $4\Delta$ rounds after it woke, its anchor is of epoch
$e_0 \geq e' - \bwin$, where $e'$ is the epoch of $r$.

\begin{algorithm}[t]
\caption{The epoch handover, run by the committee of an ending epoch, and the
witness consumer \bootstrap, run by a client with anchor
$(e_0, \comm_{e_0}, \eta_0)$.}\label{alg.handover}
\small
\begin{algorithmic}[1]
    \On{entering epoch $e$}                                             \Comment{Run by every member of $\comm_{e-1}$}
        \State $b \gets$ the block of height $h_{e-1}$
        \State cast a signature over $(e, H(b))$\label{line.hosign}
        \State $\handover_e \gets$ a quorum of $q$ such signatures by $\comm_{e-1}$\label{line.hoquorum}
        \State include $\handover_e$ in the first block above height $h_{e-1}$\label{line.hoinclude}
    \EndOn

    \Statex
    \Function{\bootstrap}{$w$}                                          \Comment{$w$ carries blocks, certificates and handovers from height $h_{e_0-1}$ on}
        \If{$w$ is not a hash chain with increasing rounds, its block of height $h_{e_0-1}$ does not hash to $\eta_0$, or a block above it is of an epoch below $e_0$}\label{line.boanchor}
            \return $\bot$                                              \Comment{Diverges before the anchor}
        \EndIf
        \For{$e' \gets e_0$ \textbf{to} the last epoch in $w$}          \Comment{$\comm_{e'}$ is known here}
            \If{some block of epoch $e'$ in $w$ has no quorum certificate of $\comm_{e'}$ or is invalid}\label{line.boverify}
                \return $\bot$                                          \Comment{Uncertified or invalid}
            \EndIf
            \State $\comm_{e'+1} \gets \committee(\sigma_{e'})$\label{line.bonext} \Comment{$\sigma_{e'}$ is the state after height $h_{e'}$}
            \State $b \gets$ the block of height $h_{e'}$ in $w$
            \If{$w$ has a block above height $h_{e'}$ and $\handover_{e'+1}$ in $w$ is not a quorum of $\comm_{e'}$ over $(e'+1, H(b))$}\label{line.bohandover}
                \return $\bot$
            \EndIf
        \EndFor
        \return the blocks of $w$ above height $h_{e_0-1}$\label{line.boreturn}
    \EndFunction
\end{algorithmic}
\end{algorithm}

\subsection{Analysis}\label{sec.dynamic.analysis}

The proofs of this subsection are in \Cref{app.dynamic}.

\begin{theorem}[Dynamic safety]\label{thm.dynamic-safety}
    The construction of \Cref{sec.construction} with $\bootstrap$ in place of
    $\Wcons$ satisfies safety (\Cref{def.safety}), whether or not the bonding
    window and client freshness hold.
\end{theorem}

The proof applies \Cref{lem.delivery}, and the anchor ledger serves as the
base of an induction over the confirmations of honest clients.

\begin{theorem}[Dynamic liveness]\label{thm.dynamic-liveness}
    Assume the bonding window, client freshness, adversarial participation and
    $\budget \leq \stake - \maxfee$. Then no honest client freezes, and the
    conclusions of \Cref{thm.liveness} hold for every honest client, for the
    ledgers that honest clients confirm.
\end{theorem}

Two conflicting witnesses that pass $\bootstrap$ first differ in an epoch
$d \geq e_0$ whose committee both designate, and the handover certificate makes
the fork an equivocation of $\comm_d$ (\Cref{lem.fork-window}), which is still
slashable, so \Cref{lem.equivocation-cost} applies.

\subsection{Necessity of Client Freshness}\label{sec.dynamic.necessity}

Client freshness cannot be dropped. The obstacle is the long-range attack of
proof-of-stake systems: a committee whose stake has been released sells
signatures on a conflicting ledger, and the adversary extends it with committees
of his own. The attack uses one property of \committee: for every set $K$ of
$3f+1$ keys and every block $p$, some valid block of epoch $e$ with parent $p$,
which \prot\ may output, makes \committee\ designate $K$ as $\comm_{e+1}$.

\noindent\textbf{The long-range attack.} Fix epochs $d$ and $e > d + \bwin$ and
a budget $\budget > 0$. Let the committees designated after the last block $b$
of epoch $d$ consist of keys that no earlier committee holds. Let an honest
client $c$ be awake during epoch $d$, fall asleep before $b$ is proposed, and
wake during epoch $e$, after step 2 below. The adversary acts as
follows.
\begin{enumerate}
    \item He follows the protocol until an always-on honest client has output a
    ledger that contains a block of epoch $d+1$ and the bonding window of
    $\comm_d$ has ended.
    \item He pays $\budget/(2q)$ to each rational member of $\comm_d$. In
    return, it signs a block $b'$ of epoch $d$ with the parent and height of
    $b$, after which \committee\ designates keys he generated, and the handover
    certificate that names $b'$.
    \item He extends $b'$ epoch by epoch up to epoch $e$, certifying every block
    and handover with keys he holds.
    \item He answers the requests of $c$ from a full node he controls, exactly
    as \fullnode\ would answer them in the execution in which $b'$ was certified
    in epoch $d$ and $b$ was produced by steps 1 to 3.
\end{enumerate}

\begin{theorem}[Necessity of client freshness]\label{thm.freshness-necessary}
    Consider any protocol in the model of \Cref{sec.model}, with epochs, a
    bonding window of $\bwin$ epochs and a function \committee\ with the
    property above, in which honest clients output only ledgers of certified
    blocks. Let $E > 7\Delta$, let the adversary hold any budget
    $\budget > 0$, and let $e > d + \bwin$. Then the protocol does not
    guarantee all of the following: safety; that every always-on honest client
    outputs a ledger with a block of epoch $d+1$ by the end of epoch $e-1$;
    and that the client $c$ above eventually does.
\end{theorem}

In the execution with the attack and in
the one where $b$ and $b'$ exchange roles, $c$ has the same local state and
receives the same messages, so it outputs the same ledger. By safety, that
ledger extends $b$ in one execution and $b'$ in the other. \fullnode\ does not
help, because $c$ cannot tell it apart from the full node of the adversary.

The two-execution argument is the classical lower bound for long-range attacks
\cite{snowwhite-full,tas-babylon,lpr-permissionless}, which assumes that old
keys become corrupt. Here rational members of old committees sell their signatures for any
positive amount, and existential honesty does not remove the bound. With
\Cref{thm.dynamic-liveness}, the threshold is exact: a client whose last
confirmed block is in epoch $e_0$ keeps the guarantees of the static case in
epoch $e$ if $e - e_0 \leq \bwin$, and not necessarily if $e - e_0 > \bwin$. Ouroboros
Genesis~\cite{genesis} avoids the bound by having honest validators erase old
keys, which a rational validator has no reason to do.

\section{Discussion}\label{sec.discussion}

\paragraph{Recovery.} A freeze is permanent: the construction has no recovery,
and recovery as in \cite{bettersafe,adversary-majority} needs an honest majority
to return. A freeze slashes at least $f+1$ validators, his own among them, and
the adversary must pay a rational one more than $\stake - \maxfee/q$
(\Cref{lem.equivocation-cost}), for instance from saved rewards.

\paragraph{Censorship.} Transactions whose fee is below a \bw\ fraction of the
available fees can be censored forever at no cost to the adversary
(\Cref{rem.liveness-limits}). Adversarial participation is the worst case for
censorship.

\paragraph{Block size and synchrony.} With bounded blocks, a reward for fuller
blocks invites padding, and the remedy is a burned fee
\cite{eip1559,tfm,tfm-foundations}; whether \Cref{thm.censorship} then holds is
open. CoBRA~\cite[Theorem 3.1]{cobra} rules out partial synchrony without
honest validators, and which weaker guarantee it admits is open.

\section{Acknowledgements}
This work was partially supported by the Austrian Science Fund (FWF)
through the SFB SpyCode project (F8512-N), and by the
Vienna Science and Technology Fund (WWTF) through the project 10.47379/ICT22045. The research of Odysseas Sofikitis was supported by Common Prefix OÜ.
\bibliographystyle{splncs04}
\bibliography{bibliography/ref}

@inproceedings{hotstuff,
author = {Yin, Maofan and Malkhi, Dahlia and Reiter, Michael K. and Gueta, Guy Golan and Abraham, Ittai},
title = {{HotStuff}: {BFT} Consensus with Linearity and Responsiveness},
year = {2019},
isbn = {9781450362177},
publisher = {Association for Computing Machinery},
address = {New York, NY, USA},
doi = {10.1145/3293611.3331591},
booktitle = {Proceedings of the 2019 ACM Symposium on Principles of Distributed Computing},
pages = {347--356},
numpages = {10},
location = {Toronto ON, Canada},
series = {PODC '19}
}

@inproceedings{pbft,
author = {Castro, Miguel and Liskov, Barbara},
title = {Practical {Byzantine} fault tolerance},
year = {1999},
isbn = {1880446391},
publisher = {USENIX Association},
address = {USA},
booktitle = {Proceedings of the Third Symposium on Operating Systems Design and Implementation},
pages = {173--186},
numpages = {14},
location = {New Orleans, Louisiana, USA},
series = {OSDI '99}
}

@misc{cobra,
      title={{CoBRA}: A Universal Strategyproof Confirmation Protocol for Quorum-based Proof-of-Stake Blockchains}, 
      author={Zeta Avarikioti and Eleftherios Kokoris Kogias and Ray Neiheiser and Christos Stefo},
      year={2025},
      eprint={2503.16783},
      archivePrefix={arXiv},
      primaryClass={cs.CR},
      url={https://arxiv.org/abs/2503.16783}, 
}

@inproceedings{adversary-majority,
  title={Consensus Under Adversary Majority Done Right},
  author={Sridhar, Srivatsan and Tas, Ertem Nusret and Neu, Joachim and Zindros, Dionysis and Tse, David},
  booktitle={Financial Cryptography and Data Security},
  series={FC '25},
  pages={108--126},
  year={2025},
  doi={10.1007/978-3-032-07035-7_7}
}

@misc{bettersafe,
      title={Better Safe than Sorry: Recovering after Adversarial Majority},
      author={Srivatsan Sridhar and Dionysis Zindros and David Tse},
      year={2023},
      eprint={2310.06338},
      archivePrefix={arXiv},
      primaryClass={cs.CR},
      url={https://arxiv.org/abs/2310.06338},
}

@inproceedings{bar,
  author = {Aiyer, Amitanand S. and Alvisi, Lorenzo and Clement, Allen and Dahlin, Mike and Martin, Jean-Philippe and Porth, Carl},
  title = {{BAR} fault tolerance for cooperative services},
  year = {2005},
  booktitle = {Proceedings of the Twentieth ACM Symposium on Operating Systems Principles},
  pages = {45--58},
  series = {SOSP '05}
}

@inproceedings{forensics,
  author = {Sheng, Peiyao and Wang, Gerui and Nayak, Kartik and Kannan, Sreeram and Viswanath, Pramod},
  title = {{BFT} Protocol Forensics},
  year = {2021},
  booktitle = {Proceedings of the 2021 ACM SIGSAC Conference on Computer and Communications Security},
  pages = {1722--1743},
  series = {CCS '21}
}

@inproceedings{polygraph,
  author = {Civit, Pierre and Gilbert, Seth and Gramoli, Vincent},
  title = {Polygraph: Accountable {Byzantine} Agreement},
  year = {2021},
  booktitle = {Proceedings of the 41st IEEE International Conference on Distributed Computing Systems},
  pages = {403--413},
  series = {ICDCS '21}
}

@inproceedings{winkle,
  author = {Azouvi, Sarah and Danezis, George and Nikolaenko, Valeria},
  title = {Winkle: Foiling Long-Range Attacks in Proof-of-Stake Systems},
  year = {2020},
  booktitle = {Proceedings of the 2nd ACM Conference on Advances in Financial Technologies},
  pages = {189--201},
  series = {AFT '20}
}

@misc{eip1559,
      title={Transaction Fee Mechanism Design for the {Ethereum} Blockchain: An Economic Analysis of {EIP-1559}},
      author={Tim Roughgarden},
      year={2020},
      eprint={2012.00854},
      archivePrefix={arXiv},
      primaryClass={cs.GT},
      url={https://arxiv.org/abs/2012.00854},
}

@inproceedings{tfm,
  title={Transaction Fee Mechanism Design},
  author={Roughgarden, Tim},
  booktitle={Proceedings of the 22nd ACM Conference on Economics and Computation},
  series={EC '21},
  year={2021},
  doi={10.1145/3465456.3467591}
}

@inproceedings{tfm-foundations,
  title={Foundations of Transaction Fee Mechanism Design},
  author={Chung, Hao and Shi, Elaine},
  booktitle={Proceedings of the 2023 ACM-SIAM Symposium on Discrete Algorithms},
  series={SODA '23},
  pages={3856--3899},
  year={2023}
}

@misc{tfm-multiproposer,
      title={Multiple Proposer Transaction Fee Mechanism Design: Robust Incentives Against Censorship and Bribery},
      author={Aikaterini-Panagiota Stouka and Julian Ma and Thomas Thiery},
      year={2025},
      eprint={2505.13751},
      archivePrefix={arXiv},
      primaryClass={cs.GT},
      url={https://arxiv.org/abs/2505.13751},
}

@misc{focil,
      title={{EIP-7805}: Fork-choice enforced Inclusion Lists {(FOCIL)}},
      author={Thomas Thiery and Francesco D'Amato and Julian Ma and Barnab\'e Monnot and Terence Tsao and Jacob Kaufmann and Jihoon Song},
      year={2024},
      howpublished={Ethereum Improvement Proposals},
      url={https://eips.ethereum.org/EIPS/eip-7805},
}

@inproceedings{rational-vs-byzantine,
  author = {Amoussou-Guenou, Yackolley and Biais, Bruno and Potop-Butucaru, Maria and Tucci-Piergiovanni, Sara},
  title = {Rational vs {Byzantine} Players in Consensus-based Blockchains},
  year = {2020},
  booktitle = {Proceedings of the 19th International Conference on Autonomous Agents and MultiAgent Systems},
  pages = {43--51},
  series = {AAMAS '20}
}

@inproceedings{rational-committee,
  author = {Amoussou-Guenou, Yackolley and Biais, Bruno and Potop-Butucaru, Maria and Tucci-Piergiovanni, Sara},
  title = {Rational Behaviors in Committee-Based Blockchains},
  year = {2021},
  booktitle = {24th International Conference on Principles of Distributed Systems (OPODIS 2020)},
  series = {LIPIcs},
  volume = {184},
  pages = {12:1--12:16},
  doi = {10.4230/LIPIcs.OPODIS.2020.12}
}

@inproceedings{tenderstake,
  title={Achieving State Machine Replication without Honest Players},
  author={McMenamin, Conor and Daza, Vanesa and Pontecorvi, Matteo},
  booktitle={Proceedings of the 3rd ACM Conference on Advances in Financial Technologies},
  series={AFT '21},
  pages={1--14},
  year={2021},
  doi={10.1145/3479722.3480986}
}

@inproceedings{trap,
  author = {Ranchal-Pedrosa, Alejandro and Gramoli, Vincent},
  title = {{TRAP}: The Bait of Rational Players to Solve {Byzantine} Consensus},
  year = {2022},
  booktitle = {Proceedings of the 2022 ACM on Asia Conference on Computer and Communications Security},
  pages = {168--181},
  doi = {10.1145/3488932.3517386},
  series = {ASIA CCS '22}
}

@misc{tendermint,
  title={The latest gossip on {BFT} consensus},
  author={Buchman, Ethan and Kwon, Jae and Milosevic, Zarko},
  year={2018},
  eprint={1807.04938},
  archivePrefix={arXiv},
  primaryClass={cs.DC},
  url={https://arxiv.org/abs/1807.04938},
}

@inproceedings{fairledger,
  title={{FairLedger}: A Fair Blockchain Protocol for Financial Institutions},
  author={Lev-Ari, Kfir and Spiegelman, Alexander and Keidar, Idit and Malkhi, Dahlia},
  booktitle={23rd International Conference on Principles of Distributed Systems (OPODIS 2019)},
  series={LIPIcs},
  volume={153},
  year={2019},
  doi={10.4230/LIPIcs.OPODIS.2019.4}
}

@inproceedings{adgh06,
  author = {Abraham, Ittai and Dolev, Danny and Gonen, Rica and Halpern, Joseph Y.},
  title = {Distributed computing meets game theory: robust mechanisms for rational secret sharing and multiparty computation},
  year = {2006},
  booktitle = {Proceedings of the Twenty-Fifth Annual ACM Symposium on Principles of Distributed Computing},
  pages = {53--62},
  series = {PODC '06}
}

@inproceedings{adh08,
  author = {Abraham, Ittai and Dolev, Danny and Halpern, Joseph Y.},
  title = {Lower Bounds on Implementing Robust and Resilient Mediators},
  year = {2008},
  booktitle = {Theory of Cryptography Conference (TCC 2008)},
  pages = {302--319},
  publisher = {Springer}
}

@article{abc,
  author = {Civit, Pierre and Gilbert, Seth and Gramoli, Vincent and Guerraoui, Rachid and Komatovic, Jovan},
  title = {As easy as {ABC}: Optimal {(A)}ccountable {(B)}yzantine {(C)}onsensus is easy!},
  journal = {Journal of Parallel and Distributed Computing},
  volume = {181},
  pages = {104743},
  year = {2023},
  doi = {10.1016/j.jpdc.2023.104743}
}

@inproceedings{accgadget,
  author = {Neu, Joachim and Tas, Ertem Nusret and Tse, David},
  title = {The Availability-Accountability Dilemma and its Resolution via Accountability Gadgets},
  year = {2022},
  booktitle = {Financial Cryptography and Data Security},
  series = {LNCS},
  volume = {13411},
  pages = {541--559},
  publisher = {Springer}
}

@inproceedings{zlb,
  author = {Ranchal-Pedrosa, Alejandro and Gramoli, Vincent},
  title = {{ZLB}: A Blockchain to Tolerate Colluding Majorities},
  year = {2024},
  booktitle = {54th Annual IEEE/IFIP International Conference on Dependable Systems and Networks (DSN)},
  pages = {209--222},
  doi = {10.1109/DSN58291.2024.00032}
}

@inproceedings{gong,
  author={Gong, Tiantian and Camilo, Gustavo Franco and Nayak, Kartik and Lewis-Pye, Andrew and Kate, Aniket},
  title={Recover from Excessive Faults in Partially-Synchronous {BFT} {SMR}},
  booktitle={34th USENIX Security Symposium},
  pages={4147--4166},
  year={2025}
}

@inproceedings{blr,
  title={Beyond Optimal Fault-Tolerance},
  author={Lewis-Pye, Andrew and Roughgarden, Tim},
  booktitle={7th Conference on Advances in Financial Technologies (AFT 2025)},
  series={LIPIcs},
  volume={354},
  pages={15:1--15:23},
  year={2025},
  doi={10.4230/LIPIcs.AFT.2025.15}
}

@inproceedings{flexbft,
  author = {Malkhi, Dahlia and Nayak, Kartik and Ren, Ling},
  title = {Flexible {Byzantine} Fault Tolerance},
  year = {2019},
  booktitle = {Proceedings of the 2019 ACM SIGSAC Conference on Computer and Communications Security},
  pages = {1041--1053},
  series = {CCS '19}
}

@inproceedings{flexopt,
  author = {Neu, Joachim and Sridhar, Srivatsan and Yang, Lei and Tse, David},
  title = {Optimal Flexible Consensus and its Application to {Ethereum}},
  year = {2024},
  booktitle = {2024 IEEE Symposium on Security and Privacy (SP)},
  pages = {3885--3903}
}

@misc{snowwhite-full,
  author={Daian, Phil and Pass, Rafael and Shi, Elaine},
  title={{Snow White}: Robustly Reconfigurable Consensus and Applications to Provably Secure Proof of Stake},
  howpublished={Cryptology ePrint Archive, Paper 2016/919, full version},
  year={2016},
  url={https://eprint.iacr.org/2016/919}
}

@inproceedings{tas-babylon,
  author={Tas, Ertem Nusret and Tse, David and Gai, Fangyu and Kannan, Sreeram and Maddah-Ali, Mohammad Ali and Yu, Fisher},
  title={Bitcoin-Enhanced Proof-of-Stake Security: Possibilities and Impossibilities},
  booktitle={IEEE Symposium on Security and Privacy},
  series={S\&P '23},
  year={2023}
}

@misc{lpr-permissionless,
  author={Lewis-Pye, Andrew and Roughgarden, Tim},
  title={Permissionless Consensus},
  year={2023},
  eprint={2304.14701},
  archivePrefix={arXiv},
  primaryClass={cs.DC},
  url={https://arxiv.org/abs/2304.14701}
}

@inproceedings{genesis,
  author={Badertscher, Christian and Ga{\v{z}}i, Peter and Kiayias, Aggelos and Russell, Alexander and Zikas, Vassilis},
  title={{Ouroboros Genesis}: Composable Proof-of-Stake Blockchains with Dynamic Availability},
  booktitle={Proceedings of the 2018 ACM SIGSAC Conference on Computer and Communications Security},
  series={CCS '18},
  pages={913--930},
  year={2018},
  doi={10.1145/3243734.3243848}
}

@misc{tendermint-lc,
  author={Braithwaite, Sean and Buchman, Ethan and Khoffi, Ismail and Konnov, Igor and Milosevic, Zarko and Ruetschi, Romain and Widder, Josef},
  title={A {Tendermint} Light Client},
  year={2020},
  eprint={2010.07031},
  archivePrefix={arXiv},
  primaryClass={cs.DC},
  url={https://arxiv.org/abs/2010.07031}
}

@inproceedings{fox-pai-resnick,
  author={Fox, Elijah and Pai, Mallesh M. and Resnick, Max},
  title={Censorship Resistance in On-Chain Auctions},
  booktitle={5th Conference on Advances in Financial Technologies (AFT 2023)},
  series={LIPIcs},
  volume={282},
  pages={19:1--19:20},
  year={2023},
  doi={10.4230/LIPIcs.AFT.2023.19}
}

@misc{price-of-censorship,
  author={Saraf, Aditya and Kaklamanis, Ioannis and Wadhwa, Sarisht and Elsheimy, Fatima},
  title={Price of Censorship: Censorship Resistance and Throughput under Rational Concurrent Proposers},
  year={2026},
  eprint={2607.16995},
  archivePrefix={arXiv},
  primaryClass={cs.GT},
  url={https://arxiv.org/abs/2607.16995}
}
\appendix
\section{Proofs of \Cref{sec.analysis}}\label{app.static}

The safety proof rests on one timing fact, which we state as a lemma.

\begin{lemma}[Delivery]\label{lem.delivery}
    Let an honest client relay a certificate at round $t$, and let an honest
    client $c$ run a confirmation at round $s \geq t + 2\Delta$. Then $c$ holds
    the certificate when the confirmation runs.
\end{lemma}
\begin{proof}
    The relay reaches \fullnode by round $t + \Delta$. Let $t' \leq t + \Delta$
    be the round at which \fullnode first receives the certificate. By
    \Cref{line.relay}, \fullnode relays it at round $t'$. Let $t_w$ be the round
    at which $c$ last woke before round $s$, or $0$ if $c$ has been awake since
    the first round. A client that falls asleep discards
    its pending confirmations, so $c$ scheduled the confirmation at a round
    $t'' \geq t_w$ and is awake from round $t_w$ to round $s$. By
    \Cref{line.cschedule}, $s = t'' + 2\Delta \geq t_w + 2\Delta$.

    Case 1: $t_w \leq t'$. Then $c$ is awake from round $t'$ to round
    $t' + \Delta \leq t + 2\Delta \leq s$, and the relay of \fullnode reaches it
    by round $t' + \Delta$.

    Case 2: $t_w > t'$. At round $t_w$, $c$ multicasts a request
    (\Cref{line.crequest}). The request reaches \fullnode by round
    $t_w + \Delta$. \fullnode holds the certificate from round $t' < t_w$, so the
    reply of \Cref{line.serve} carries it and reaches $c$ by round
    $t_w + 2\Delta \leq s$.

    In both cases $c$ receives the certificate by round $s$ and keeps it in $R$,
    since it does not sleep between rounds $t_w$ and $s$. A client processes the
    messages of a round before its confirmations, so $c$ holds the certificate
    when the confirmation runs.
\end{proof}

\begin{proofof}{\Cref{thm.safety}}
    Suppose towards a contradiction that honest clients $c_1$ and $c_2$ output
    ledgers $\ledger_1 = \ledger^{c_1}_{r_1}$ and $\ledger_2 =
    \ledger^{c_2}_{r_2}$ with $\ledger_1 \not\sim \ledger_2$. By
    \Cref{line.cextend} the output of a client only grows by extension, so
    $c_1$ confirmed $\ledger_1$ at some round $s_1 \leq r_1$ and $c_2$ confirmed
    $\ledger_2$ at some round $s_2 \leq r_2$. By symmetry we may assume
    $s_1 \leq s_2$.

    Let $t_1$ be the round at which the witness for $\ledger_1$ became complete
    at $c_1$. By \Cref{line.cschedule}, $s_1 = t_1 + 2\Delta$. By
    \Cref{line.cgossip}, $c_1$ relayed every certificate of the witness by
    round $t_1$. By \Cref{lem.delivery}, $c_2$ holds all of them when its
    confirmation at round $s_2 \geq t_1 + 2\Delta$ runs. The consumer \Wcons is
    deterministic, so by \Cref{line.cwitness,line.cseen} $c_2$ has added
    $\ledger_1$ to \seen\ by then. Then \Cref{line.cconflict} stops $c_2$ from
    confirming $\ledger_2$ at round $s_2$. Contradiction.
\end{proofof}

\begin{proofof}{\Cref{lem.equivocation-cost}}
    Let $\qc_b$ and $\qc_{b'}$ certify conflicting blocks $b$ and $b'$. Each
    carries $q = 2f+1$ signatures of distinct validators out of $n = 3f+1$, so
    $|\signers(\qc_b) \cap \signers(\qc_{b'})| \geq 2q - n = f + 1$. Every
    validator in the intersection equivocates. At most $f$ validators are
    Byzantine, so the intersection contains a rational validator.

    Let $v$ be a rational validator that loses its stake \stake\ for an
    equivocation.
    By \Cref{thm.safety,thm.dynamic-safety}, no two honest clients confirm
    conflicting ledgers. So $v$ receives at most one block reward from the
    equivocation, which is at most $\maxfee/q$, and gains nothing from a
    double-spend. Had $v$ not equivocated, its utility would be non-negative. A
    rational validator deviates only for strictly greater utility, and a side
    payment is made before the action it buys. So the adversary pays $v$ more
    than $\stake - \maxfee/q$ before $v$ signs. Let $h$ be the height to which
    this payment is charged; a payment for a signature on $\handover_e$ is
    charged to the height of the block that $\handover_e$ names. The adversary
    receives $g_h \leq \frac{f}{q}\maxfee$ from height $h$, since the block
    rewarded at height $h$ is on a ledger that an honest client confirms, whose
    committees satisfy the bound on Byzantine members, in the dynamic setting by
    item 2 of \Cref{lem.fork-window}. Hence
    $\budget_h > \stake - \frac{1}{q}\maxfee - \frac{f}{q}\maxfee \geq
    \stake - \maxfee$, which is positive since $\stake > \maxfee$.

    By accountability, the rational validator in the intersection loses its
    stake, so the adversary draws more than $\stake - \maxfee$ from \budget. If
    $\budget \leq \stake - \maxfee$, this contradicts
    $\sum_h \budget_h \leq \budget$, so no two conflicting blocks are certified.
    A client freezes only when \seen\ holds two conflicting ledgers
    (\Cref{line.cconflict}). These contain conflicting blocks at the first
    position where they differ, and a witness certifies every block of its
    ledger, so no honest client freezes.
\end{proofof}

\begin{proofof}{\Cref{thm.censorship}}
    Let $b$ be the block certified at height $h$, proposed at round $r$, and
    let $\mathcal{R}$ be the set of rational signers of $\qc_b$. Write
    $V = \reward(\mempool_h)$ and $\cens = \cens_h$. Then
    $|\mathcal{R}| = q - f^{*} = (1-\bw)q$ and $\reward(b) = V - \cens$.

    Upper bound. Let $v \in \mathcal{R}$. By adversarial participation, $v$ compares
    signing $b$ with signing $b \cup \mempool^v_r$. By
    \Cref{def.censored}, every transaction of $\mempool_h$ that is not in $b$
    is held by $v$, so the second block carries at least \cens\ more fees and
    pays $v$ at least $\cens/q$ more. Fees on the adversary's own transactions
    appear in both blocks, so they do not change this difference. Neither
    alternative involves slashing or a double-spend. Since $\cens > 0$, $v$
    signs $b$ only if the adversary pays it more than $\cens/q$ before it
    signs. The side payments charged to height $h$ therefore exceed
    $(1-\bw)\cens$. His validators receive $g_h \leq \bw\reward(b) = \bw(V - \cens)$ from
    $b$. By the definition of $\budget_h$,
    \[
        (1-\bw)\cens < \bw(V - \cens) + \budget_h \,,
    \]
    and adding $\bw\cens$ to both sides gives $\cens < \bw V + \budget_h$.

    Tightness. Let every $v \in \mathcal{R}$ hold the same set $M$ of pending
    transactions, and let $X \subseteq M$ have total fee $c > 0$. The
    adversary has $b = M \setminus X$ proposed, by one of his validators or by
    a paid rational signer, signs it with $f^{*} = \bw q$ of
    his validators, and pays each rational signer $(c + \varepsilon)/q$. A
    rational signer receives $(V - c)/q + (c + \varepsilon)/q$ by signing $b$
    and $V/q$ by signing $b \cup M = M$, so it strictly prefers to sign $b$.
    The block $b$ is certified with $\mempool_h = M$ and $\cens_h = c$. The
    adversary pays $(1-\bw)(c + \varepsilon)$ and receives
    $g_h = \bw(V - c)$ once $b$ is confirmed, so
    $\budget_h = \max(0,\, c - \bw V + (1-\bw)\varepsilon) \leq
    \max(0,\, c - \bw V + \varepsilon)$.
\end{proofof}

\begin{proofof}{\Cref{cor.inclusion}}
    Suppose towards a contradiction that $\tx \notin b$. Then $\tx \in \mempool_h \setminus b$, so
    $\cens_h \geq \tx.\fee > 0$, and \Cref{thm.censorship} gives
    $\tx.\fee \leq \cens_h < \bw\reward(\mempool_h) + \budget_h$. Substituting
    $\reward(\mempool_h) = \tx.\fee + \reward(\mempool_h \setminus \{\tx\})$
    and rearranging gives
    $(1-\bw)\tx.\fee < \bw\reward(\mempool_h \setminus \{\tx\}) + \budget_h$,
    which contradicts the hypothesis. For the second form, the maps
    $\bw \mapsto \frac{\bw}{1-\bw}$ and $\bw \mapsto \frac{1}{1-\bw}$ are
    increasing on $[0,1)$, so the right-hand side is largest at
    $\bw = \frac{f}{2f+1}$, where it equals
    $\frac{f}{f+1}\reward(\mempool_h \setminus \{\tx\}) +
    \frac{2f+1}{f+1}\budget_h$.
\end{proofof}

\begin{proofof}{\Cref{thm.liveness}}
    By \Cref{lem.equivocation-cost}, no two conflicting blocks are certified
    and no honest client freezes.

    Item 1. Let $c$ be awake throughout $[s', s' + 4\Delta]$ and let $t_w \leq s'$
    be the round at which it last woke. Consider a certificate of \ledger\ and
    let $t' \leq s$ be the round at which \fullnode first received it. If
    $t' \geq t_w$, the relay of \fullnode at round $t'$ (\Cref{line.relay})
    reaches $c$ by round $t' + \Delta \leq s' + \Delta$. If $t' < t_w$, the
    request of $c$ at round $t_w$ (\Cref{line.crequest}) reaches \fullnode by
    round $t_w + \Delta$, and the reply of \Cref{line.serve} carries the
    certificate and reaches $c$ by round $t_w + 2\Delta \leq s' + 2\Delta$. So
    the witness for \ledger\ is complete at $c$ by round $s' + 2\Delta$, and
    by \Cref{line.cschedule} its confirmation runs by round $s' + 4\Delta$.
    Every ledger in \seen\ and the output \out\ consist of certified blocks,
    and no two conflicting blocks are certified. So \Cref{line.cconflict}
    passes, and \out\ is consistent with \ledger. If $\out \preceq \ledger$,
    then \Cref{line.coutput} outputs \ledger. Otherwise $\ledger \preceq \out$.
    In both cases the output extends \ledger.

    Item 2. The submitting client sends \tx\ to every validator, and
    validators are always awake, so from round
    $r + \Delta$ every validator holds \tx\ until it is included. Let $h$ be a
    height whose block $b$ is proposed after round $r + \Delta$ and certified
    with bandwidth \bw, and let $\tx \notin b$. Then $\tx \in \mempool_h$. By
    \Cref{cor.inclusion},
    \[
        \budget_h > (1-\bw)\tx.\fee - \bw\reward(\mempool_h \setminus \{\tx\})
        = (1-\bw)\Bigl(\tx.\fee - \frac{\bw}{1-\bw}
          \reward(\mempool_h \setminus \{\tx\})\Bigr) .
    \]
    Since $\frac{\bw}{1-\bw} \leq \frac{f}{f+1}$, the bracket is at least
    $\delta$. Since $1 - \bw \geq \frac{f+1}{2f+1} > \frac{1}{2}$, we get
    $\budget_h > \delta/2$. As $\sum_h \budget_h \leq \budget$, there are at
    most $2\budget/\delta$ such heights before \tx\ is included.
\end{proofof}

\section{Proofs of \Cref{sec.dynamic}}\label{app.dynamic}

\begin{proofof}{\Cref{thm.dynamic-safety}}
    The proof is by a minimal counterexample over the confirmations of honest
    clients, ordered by round with ties broken arbitrarily. It uses two facts
    about a ledger \ledger\ that an honest client confirms at round $s$. First,
    every certificate of \ledger\ was relayed by an honest client by round
    $s - 2\Delta$. Second, the rounds of the blocks of \ledger\ increase, and every block
    of \ledger\ is valid and carries a quorum certificate of the committee that \ledger\ designates for its epoch.
    Every handover certificate in \ledger\ is a quorum of the committee of
    the epoch it ends. Both facts hold by induction on $s$. Let $a$ be the
    ledger from which the client took the anchor $(e_0, \comm_{e_0}, \eta_0)$
    that it used for \ledger. The blocks, certificates and handover certificates
    of \ledger\ from height $h_{e_0-1}$ onward form the witness, which the client relayed at \Cref{line.cgossip} by the
    round $t$ at which it became complete. By \Cref{line.cschedule},
    $s = t + 2\Delta$. \bootstrap\ checked them at
    \Cref{line.boverify,line.bohandover}. The output of the client starts at
    $a$ and grows by extension (\Cref{line.cextend}), so $a \preceq \ledger$,
    and the blocks and handover certificates of \ledger\ up to the last block of
    $a$ belong to $a$. This covers $\handover_{e_0}$, which lies in the first
    block above height $h_{e_0-1}$. Either $a$ is empty, or the client confirmed
    $a$ at an earlier round and the induction hypothesis applies.

    Towards a contradiction, let the confirmation of $\ledger_2$ by $c_2$ at
    round $s_2$ be the first that conflicts with an earlier one, and let $c_1$
    have confirmed $\ledger_1 \not\sim \ledger_2$ earlier, at round
    $s_1 \leq s_2$. By the first fact, every certificate of $\ledger_1$ was
    relayed by an honest client by round $s_1 - 2\Delta \leq s_2 - 2\Delta$.
    By \Cref{lem.delivery}, $c_2$ holds all of them when its confirmation at
    round $s_2$ runs. Let $a$ be the ledger from which $c_2$ took the anchor it
    holds when it receives the last of them. Either $a$ is empty or $c_2$
    confirmed it before round $s_2$, so $a \sim \ledger_1$ by the choice of
    $\ledger_2$. The output of $c_2$ grows only by extension
    (\Cref{line.cextend}) and $c_2$ outputs $\ledger_2$ at round $s_2$, so
    $a \preceq \ledger_2$. If $\ledger_1 \preceq a$, then
    $\ledger_1 \preceq \ledger_2$, which contradicts
    $\ledger_1 \not\sim \ledger_2$. Hence $a \preceq \ledger_1$.

    The blocks, certificates and handover certificates of $\ledger_1$ from
    height $h_{e_0-1}$ onward form a witness. It passes \Cref{line.boanchor}
    because $a \preceq \ledger_1$: the rounds of $\ledger_1$ increase by the
    second fact, and its block after the anchor block is that of $a$, which is
    of epoch $e_0$. The ledgers $a$ and $\ledger_1$ agree up to
    height $h_{e_0-1}$, so they designate the same committee $\comm_{e_0}$,
    and by the second fact the witness passes
    \Cref{line.boverify,line.bohandover}. So $\bootstrap$ returns the blocks of
    $\ledger_1$ above height $h_{e_0-1}$, and $c_2$ adds $\ledger_1$ to $\seen$
    at \Cref{line.cseen} before its confirmation runs. Then
    \Cref{line.cconflict} stops $c_2$ from confirming $\ledger_2$ at round
    $s_2$, which contradicts the choice of $\ledger_2$.
\end{proofof}

The next lemma shows that, for a fresh client, the bound on Byzantine members
covers every committee that matters, and that a fork it accepts is an
equivocation of one such committee.

\begin{lemma}[Fork epoch]\label{lem.fork-window}
    Assume client freshness. Let an honest client be awake at round $x$ with an
    anchor of epoch $e_0$ taken from a ledger $a$, and call a ledger
    \emph{admissible} if it agrees with $a$ up to height $h_{e_0-1}$ and passes
    the checks of $\bootstrap$ above that height.
    \begin{enumerate}
        \item Round $x$ is at most $4\Delta$ rounds after the end of epoch
        $e_0 + \bwin$.
        \item The bound on Byzantine members holds for every committee that an
        admissible ledger, or a ledger that an honest client confirms,
        designates.
        \item If two admissible ledgers conflict, then there is an epoch
        $d \geq e_0$ such that at least $f+1$ members of $\comm_d$ equivocated
        on the blocks of epoch $d$ or on $\handover_{d+1}$, and at least one of
        them is rational.
    \end{enumerate}
\end{lemma}
\begin{proof}
    Item 1. The anchor of a client only moves forward. If $x$ comes at least
    $4\Delta$ rounds after the client woke, then by client freshness $x$ lies in
    an epoch at most $e_0 + \bwin$. Otherwise the client woke fewer than
    $4\Delta$ rounds before $x$, during an epoch $e$ with
    $e - \bwin \leq e_0$, so $x$ is less than $4\Delta$ rounds after the end of
    epoch $e_0 + \bwin$.

    Item 2. Let \ledger\ be admissible. By \Cref{line.boanchor}, the blocks of
    \ledger\ above height $h_{e_0-1}$ are of epochs at least $e_0$, and the
    client holds their certificates by round $x$. By item 1, each was formed
    within the bonding window of its signers, which lasts until $5\Delta$
    rounds after the end of an epoch at least $e_0 + \bwin$. The blocks at or
    below height $h_{e_0-1}$ are those of $a$. The client confirmed $a$ at an
    earlier round, while fresh, so the same argument applies to $a$, and by
    induction over its confirmations down to the genesis block. The same
    induction applies to every ledger that an honest client confirms.

    Item 3. Let \ledger\ and $\ledger'$ be the two ledgers. They agree up to
    height $h_{e_0-1}$. Let $i$ be the first position at which they differ, and
    let $y$ and $y'$ be their blocks at position $i$. Both have the same parent
    $p$. Let $d \geq e_0$ be the smaller of the epochs of $y$ and $y'$, say the
    epoch of $y$. The rounds of the blocks of a ledger increase with height, so
    every block of an epoch before $d$ lies at or below $p$ in both ledgers.
    Hence both ledgers have the same state $\sigma_{d-1}$ and designate the
    same committee $\comm_d$, to which the bound applies by item 2.

    Case 1: $y'$ is also of epoch $d$. By \Cref{line.boverify}, $y$ and $y'$
    carry quorum certificates of $\comm_d$. They conflict, since they differ
    and have the same parent.

    Case 2: $y'$ is of a later epoch. Then $p$ is the block of height $h_d$ in
    $\ledger'$, and $\ledger'$ has a block above it. By
    \Cref{line.bohandover}, $\ledger'$ carries $\handover_{d+1}$, a quorum of
    $\comm_d$ over $(d+1, H(p))$. By \Cref{line.boverify}, $y$ carries a quorum
    certificate of $\comm_d$, and $y$ is a block of epoch $d$ that extends $p$.

    In both cases two quorums of $\comm_d$ sign an equivocation. They share at
    least $2q - n = f+1$ members, and at most $f$ members of $\comm_d$ are
    Byzantine.
\end{proof}

\begin{proofof}{\Cref{thm.dynamic-liveness}}
    Suppose towards a contradiction that an honest client $c$ freezes. Then
    \seen\ holds two conflicting ledgers $\ledger_1$ and $\ledger_2$, and let
    $\ledger_2$ be the later to enter \seen, at round $x$. Let $a$ be the
    ledger of the anchor of $c$ at round $x$, of epoch $e_0$. The ledger
    $\ledger_2$ passed $\bootstrap$
    under this anchor, so it is admissible. Either $\ledger_1$ also passed
    $\bootstrap$ under it, or $c$ confirmed $a$ while $\ledger_1$ was in \seen,
    in which case $\ledger_1 \sim a$ by \Cref{line.cconflict}. Since
    $\ledger_1$ conflicts with $\ledger_2$, which agrees with $a$ up to height
    $h_{e_0-1}$, $\ledger_1$ also agrees with $a$ up to that height. Its blocks
    above it either belong to $a$, which an honest client confirmed, or were
    checked by $\bootstrap$ with the committees that its own prefix designates,
    so $\ledger_1$ is admissible as well.

    By item 3 of \Cref{lem.fork-window}, a rational member $v$ of some
    $\comm_d$ with $d \geq e_0$ equivocated on the blocks of epoch $d$ or on
    $\handover_{d+1}$. The client relays both witnesses on receipt to the full
    nodes and the validators (\Cref{line.cgossip}), so the committee of the
    current epoch holds the evidence by round $x + \Delta$. By item 1 of
    \Cref{lem.fork-window}, this is at most $5\Delta$ rounds after the end of
    epoch $e_0 + \bwin \leq d + \bwin$, inside the bonding window of
    $\comm_d$, so $v$ loses its stake \stake. By \Cref{lem.equivocation-cost}, the
    adversary draws more than $\stake - \maxfee$ from \budget, which
    contradicts $\budget \leq \stake - \maxfee$. So no honest client freezes.

    Item 1 of \Cref{thm.liveness} concerns a ledger \ledger\ that an honest
    client has confirmed. \fullnode\ holds its certificates and handover
    certificates. By \Cref{thm.dynamic-safety}, the ledger $a$ of the anchor of
    $c$ satisfies $a \sim \ledger$. If $\ledger \preceq a$, the output of $c$
    already extends \ledger. Otherwise $a \preceq \ledger$, and the witness of
    \ledger\ from height $h_{e_0-1}$ passes $\bootstrap$, since \ledger\ was
    confirmed by an honest client and its blocks above $a$ are of epochs at
    least $e_0$. The timing of the static proof applies, $c$ does not freeze,
    and $\out \sim \ledger$ by \Cref{thm.dynamic-safety}, so the output of $c$
    extends \ledger. Item 2 concerns the heights of ledgers that honest clients
    confirm. By item 2 of \Cref{lem.fork-window}, their committees satisfy the
    bound on Byzantine members, so $\bw \leq f/q$ at each such height, and
    \Cref{thm.censorship,cor.inclusion} apply as in the static case.
\end{proofof}

\begin{proofof}{\Cref{thm.freshness-necessary}}
    Suppose the protocol guarantees safety and that every always-on honest
    client eventually outputs a ledger that contains a block of epoch $d+1$. We
    show that $c$ outputs no such ledger in two executions A and B.

    Execution A. The adversary runs the long-range attack. By the second
    guarantee, an always-on honest client outputs a ledger $\ledger_A$ that
    contains a block of epoch $d+1$ by the end of epoch $e-1$. The bonding
    window of $\comm_d$ ends $5\Delta$ rounds after the end of epoch
    $d + \bwin \leq e - 1$, and buying the signatures of step 2 takes at most
    $2\Delta$ rounds more. Since $E > 7\Delta$, steps 1 and 2 end within epoch
    $e$, and $c$ can wake in epoch $e$ after them. Let $b$ be the last block
    of epoch $d$ in $\ledger_A$ and let $P$ be the prefix of $\ledger_A$ below
    $b$. In step 2, the bonding window of $\comm_d$ has ended, and the keys
    designated after $b$ are fresh, so a member of $\comm_d$ holds no stake and
    no seat that signing $b'$ and its handover certificate could affect.
    Refusing gains it nothing, so a rational member strictly prefers the
    payment and signs. The Byzantine members sign as well. The adversary pays at
    most $(2f+1)\budget/(2q) = \budget/2$. The branch through $b'$ is certified
    after the bonding window of $\comm_d$, so the bound on Byzantine members
    does not apply to the committees it designates, and the adversary may hold
    their keys. After step 3, every block and handover certificate on the
    branch through $b'$ carries a quorum of the committee that its own prefix
    designates, as on the branch through $b$.

    Execution B. By the property of \committee, some valid block $b'$ of epoch
    $d$ with the parent of $b$, which \prot\ may output, designates a set of
    fresh keys. Let B agree with A until $b$ is proposed. In B the protocol
    outputs $b'$ as the last block of epoch $d$, and the keys that the branch
    through $b'$ designates belong to validators, $f$ Byzantine and $2f+1$
    rational in every committee. The adversary runs steps 1 to 4 with $b$ in
    place of $b'$: he buys the signatures of $\comm_d$ on $b$ and its handover
    certificate, and holds the fresh keys that the branch through $b$
    designates, to which the bound again does not apply. By the second
    guarantee, the always-on honest client of B outputs a ledger $\ledger_B$
    that contains a block of epoch $d+1$ by the end of epoch $e-1$, before
    step 2 of B. Honest clients
    output only ledgers of certified blocks, and until that step the only
    certified block with the parent and height of $b$ is $b'$. So $\ledger_B$
    extends $P$ by $b'$.

    Fix the random coins of $c$ to be the same in A and B. The two executions
    agree until $b$ is proposed, and $c$ sleeps from then until it wakes in
    epoch $e$, so $c$ has the same local state in both. After waking, $c$
    receives messages from full nodes and validators only. In A, the adversary
    sends to $c$, from his full node and from the keys of the branch through
    $b'$, exactly the messages that \fullnode\ and the validators of the branch
    through $b'$ send to $c$ in B. In B he does the same with $b$ and $b'$
    swapped. He delivers every message to $c$ exactly $\Delta$ rounds after it
    is sent. So $c$ receives
    the same messages at the same rounds in A and B, and outputs the same
    ledgers.

    Suppose towards a contradiction that $c$ outputs a ledger $\ledger_c$ that
    contains a block of epoch $d+1$. In A, safety gives
    $\ledger_c \sim \ledger_A$. The blocks of $P$ are of epoch at most $d$, so
    $\ledger_c$ is not a prefix of $P$, and $\ledger_c$ agrees with $\ledger_A$
    at the position of $b$. The block of $\ledger_c$ at that position is
    therefore $b$. In B, the same
    argument with $\ledger_B$ gives $b'$. This contradicts $b \neq b'$. So if
    safety holds in A and in B, then $c$ outputs no ledger that contains a
    block of epoch $d+1$, in either execution.
\end{proofof}

\end{document}